\documentclass[12pt]{article} 

\usepackage{amsfonts}
 \usepackage{amssymb}
 \usepackage{amsmath}

\def\pmx{\begin{pmatrix}}
\def\emx{\end{pmatrix}}
\def\bsq{\begin{subequations}}
\def\esq{\end{subequations}}
\newtheorem{thm}{Theorem}
\newtheorem{lemma}[thm]{Lemma}

\def\be{\begin{eqnarray}}
\def\ee{\end{eqnarray}}
\def\bee{\begin{eqnarray*}}
\def\eee{\end{eqnarray*}}
\def\ds{\displaystyle}
\def\bra{\langle}
\def\ket{\rangle}
\def\dg{\dagger}

\def\iff{\Leftrightarrow}

\def\raw{\rightarrow}
\def\Raw{\Rightarrow}
\def\half{{\textstyle \frac{1}{2}}}

 \def\tr{\hbox{Tr} \,}
\def\mm{ \! - \! }
\def\ts{\textstyle}
\def\nn{\nonumber}
\def\ot{\otimes}
\def\wtd{\widetilde}
\def\ovb{\overline}

\title{Lieb's simple proof of concavity of $(A,B) \mapsto  \tr A^p K^{\dg} B^{1-p} K$
\\ and remarks on related inequalities}

\author{Mary Beth Ruskai  \thanks{Partially supported  by
 the National Security Agency (NSA) and
 Advanced Research and Development Activity (ARDA) under
Army Research Office (ARO) contract number 
     DAAD19-02-1-0065, and by the National Science
        Foundation under Grant  DMS-0314228.}
\\ Department of Mathematics, Tufts University \\  Medford, Massachusetts  02155 USA
\\  {\small  marybeth.ruskai@tufts.edu} }

\begin{document}
\maketitle

\begin{abstract}
A simple, self-contained  proof is presented for the concavity of 
the map $(A,B) \mapsto  \tr A^p K^{\dg} B^{1-p} K$.    The author makes no
claim to originality; this note gives Lieb's original argument 
in its simplest, rather than its most general, form.  A sketch of the
chain of implications from this result to concavity of 
$A \mapsto \tr e^{K + \log A}$
is then presented.   An independent elementary proof is given for the joint 
convexity of the map 
$(A,B,X)  \mapsto       \tr \int_0^{\infty} X^{\dg}  \frac{1}{A+ uI} X \frac{1}{B+ uI} du$ which plays
      a key role in entropy inequalities.
\end{abstract}

\section{Introduction}

Properties of quantum entropy, particularly the inequality
known as strong subadditivity (SSA), play an important role in
quantum information theory.     The original proof of SSA is
based on a concavity result of Lieb, and several approaches
to SSA use either this result or one of his related convex trace
functions \cite{Lb}.     These results have acquired an
undeserved reputation as difficult to prove; indeed, the influential
 book   by  Nielsen and Chuang \cite{NC}
states on p. 645  that ``no transparent proof of SSA is known."
This note is intended to remedy this situation.

During recent lectures, I presented Lieb's original proof
of the joint concavity of the map $(A,B)  \mapsto \tr A^p K^{\dg} B^{1-p} K$
and was reminded just how simple and elegant it really is.
Unfortunately, some of this simplicity was lost when published \cite{Lb}
in a form that lent itself to generalizations, such as the concavity
of $(A,B) \mapsto  \tr A^p K^{\dg} B^q K$ with $p + q \leq 1$.
In view of the renewed interest in this result,
 and   frequent reference to the elementary, but long, 
 proof\footnote{Based on an argument of Uhlmann \cite{Uhl1} as 
 presented by  Simon \cite{S} and Wehrl \cite{W}.}
given in an Appendix to Nielsen and Chuang \cite{NC}, it seemed
worth making a simple argument available to a larger audience.  

 Lieb's results in \cite{Lb} include the following three theorems which we
 consider only for finite dimensional matrices.
\begin{thm}  \label{thm1}
Let $K$  be a fixed $n \times n$ matrix.  For $0 \leq p \leq 1$,   the map 
$(A,B)  \mapsto   F(A,B) \equiv   \tr A^p K^{\dg} B^{1-p} K$ 
 is well-defined on the cone of
pairs of positive, semi-definite $n \times n$ matrices $A,B$.
Moreover, for each $p  \in (0,1)$, the function
$F(A,B)$ is a concave map from this cone to  $[0,\infty)$.
\end{thm}
\begin{thm}   \label{thm2}
 The map 
$(A,B,K)  \mapsto   F(A,B,K) \equiv 
      \tr \int_0^{\infty} K^{\dg} \frac{1}{A+ uI} K \frac{1}{B+ uI} du $
 is well-defined when $A, B$ are  
positive, semi-definite $n \times n$ matrices, $\ker(A) \subset \ker(K)$
and $\ker(A) \subset \ker(K^{\dg})$.
Moreover,  $F(A,B,K)$  is jointly convex in $A,B,K$.
\end{thm}
\begin{thm}  \label{thm3}
Let $K$  be a fixed, self-adjoint $n \times n$ matrix.
The map $A \mapsto F(A) \equiv \tr e^{K + \log(A)}$ is well-defined
on the cone of positive definite matrices.  Moreover, $F(A)$ is 
concave on this cone.
\end{thm}

Lieb's proof of Theorem~\ref{thm1} uses the fact that the modulus of a function 
which is analytic and uniformly bounded
 on a strip is bounded by its supremum on the boundary.   In order to make
 this note self-contained and accessible to readers with varied background,
 we  explain this result and sketch a proof   in Appendix~\ref{app:maxmod}.
   The three  functions in the  theorems above satisfy a 
 homogeneity condition  of the form $F(\lambda A) = \lambda F(A)$ or
$ F(\lambda A,\lambda B)= \lambda  F(A,B)$, etc.   which
 has useful consequences summarized  in Appendix~\ref{app:homo}.

 In an earlier review \cite{Rusk} the author emphasized the role of the related
 concavity of $A \mapsto \tr e^{K + \log A}$, and presented Epstein's proof \cite{E}
 of this result.    In fact, both Lieb's and Epstein's methods can be used to prove
 a much larger class of inequalities whose equivalence was established by
 Lieb in \cite{Lb}.    Epstein gives a more direct route to concavity
 of $A \mapsto \tr e^{K + \log A}$; while Lieb's is simpler for 
 $(A,B)  \mapsto \tr A^p K^{\dg} B^{1-p} K$.  
 Epstein's approach  requires some deep  results from complex analysis, 
 while Lieb's argument requires  only  the maximum modulus principle.
 However, in the generalization to $\tr A^p K^{\dg} B^{q} K$, Lieb also uses a result
 about concave operator functions.   The theory of monotone and convex
 operator functions is closely connected to Epstein's approach and 
has other applications in quantum information theory.   
(For some examples and references, see \cite{LesR}.)
  
 In \cite{Rusk}, the author   showed how to use the concavity of 
 $\tr e^{K + \log A}$   to obtain simple proofs of the strong subadditivity 
 of quantum entropy and   the joint convexity of
 relative entropy.     The advantage to the
presentation in  \cite{Rusk}  is the ease with which equality conditions are
obtained.    However, the proof of concavity of 
$(A,B)  \mapsto \tr A^p K^{\dg} B^{1-p} K$ given here also yields a
short route to the same group of entropy inequalites. 
  By observing that
  \be
    \lim_{p \raw 0} \, \tfrac{1}{p} \big(\tr A^{1-p}  B^{p} - \tr A \big)
 = -\tr A (\log A - \log B)
 \ee
   one can see that Theorem~\ref{thm1} also yields a simple
proof (first noted in \cite{Lind74}) of the  joint convexity of  relative entropy. 

In section 3, Lieb's arguments showing a series of implications from
Theorem~\ref{thm1}  to Theorems \ref{thm2}  and  \ref{thm3} 
are sketched.      The goal is only to give the reader a feel for the
ideas connecting seemingly disparate results.    For more details, the
presentation in Chapter 3 of \cite{OP} is recommended.

In Section~\ref{sect:newpf}, a simple new proof of Theorem~\ref{thm2} 
is presented, together with some remarks about its connection to entropy
inequalities.  

\section{Lieb's proof of Theorem~\ref{thm1}}  \label{sect:Lb}

Theorem \ref{thm1} will first be proved for the special case $A = B$.   
Concavity then means that for each
pair $\lambda_1, \lambda_2 > 0$ with $\lambda_1 + \lambda_2 = 1$,
\be  \label{eq:main}
 \lambda_1\tr A_1^p K^{\dg} A_1^{1-p} K + \lambda_2   \tr A_2^p K^{\dg} A_2^{1-p} K  
     \leq  \tr C^p K^{\dg} C^{1-p} K 
\ee
where $C =  \lambda_1A_1 + \lambda_2   A_2$.    Observe that 
$\bra  \phi , C \phi \ket  = 0 \Raw  \bra \phi  , A_1 \phi \ket =  \bra \phi , A_2 \phi \ket = 0$ so that
(\ref{eq:main}) holds trivially for $\phi \in \ker(C)$.  Hence, it suffices to prove
the inequality on  $\ker(C)^{\perp}$ so that we can assume without loss of
generality that $C$ is invertible.  In finite dimensions this also
implies that  $C^{-1} $ is bounded.  Now define $M = C^{(1-p)/2} K C^{p/2}$
and
\be
   f_k(p) = 
      \tr A_k^p C^{-p/2} M^{\dg} C^{-(1-p)/2} A_k^{1-p} C^{-(1-p)/2} M C^{-p/2} \qquad k =1,2.
\ee
Then (\ref{eq:main}) is equivalent to
\be  \label{eq:Mbd}
 f(p) \equiv    \lambda_1f_1(p)  + \lambda_2 f_2(p) \leq \tr M^{\dg} M .
\ee
Observe that the functions above can be analytically continued to the strip
$0 \leq  {\rm Re} \,(z) \leq 1$.   

The next step is to show that each $f_k(z)$ is  bounded on
this strip.   To do this, write $z = x + i y$  with $x,y$ real, and
observe that for $A > 0$, $ \|A^{iy}\|  = 1$ and 
$ \|A^{x}\|  = \sup_{\| \psi \| = 1} \bra \psi, A^x \psi \ket =
(\sup_{\| \psi \| = 1} \bra \psi, A \psi \ket)^x = \| A \|^x$ is the (largest eigenvalue of $A)^x$.
Then by repeated application of the inequality
$|\tr XY| \leq  \tr |XY| \leq \|X\|_{\infty} \tr |Y|$ (where  $\|X\|_{\infty} \equiv \| X \|$
is the usual operator sup norm as above),   one finds that
\be  \label{unifbnd}
  |f_k(z))| \leq  \| A_k \| \, \| C^{-1} \| \, \tr M^{\dg} M \leq 
       \| A_k \| \, \| C^{-1} \| \, \|C \| \,  \tr K^{\dg} K.
\ee
for  $0 \leq x \leq 1$.
Thus  the functions $f_k(z)$ are uniformly bounded on the strip $0 \leq  {\rm Re} \,(z) \leq 1$. 

By the maximum modulus principle (Appendix~\ref{app:maxmod}),
 $|f_k(z)| $ is bounded by its supremum on
the boundary of this strip, i.e., for $z = 0 +iy$ or $z = 1 +iy$.  Now, 
\be   \label{eq6}
\lefteqn{  f_k(0+iy)}   \\  & =  & \tr \big( A_k^{iy/2} C^{-iy/2} 
  M^{\dg} C^{iy/2} C^{-1/2} A_k^{1/2}  \big) 
      \big( A_k^{-iy}  A_k^{1/2}  C^{-1/2}  C^{iy/2} M C^{-iy/2} A_k^{iy/2} \big) \nonumber
  \ee 
 has the form  $\tr X^{\dg} Y$ which is bounded above 
 by $ \big( \tr X^{\dg} X  \, \tr Y^{\dg} Y \big)^{1/2}$.
 Since operators of the form $A^{it}$ are unitary for $t$ real
 and $A$ positive,   one finds
  \be  \label{eq7}
   | f_k(0+iy)| \leq  \tr M^{\dg} C^{iy/2} C^{-1/2} A_k C^{-1/2} C^{-iy/2} M \qquad k=1,2 .
 \ee 
Thus
\be  \label{eq8}
  |f(0+iy) | & \leq  &  \lambda_1|f_1(0+iy) | + \lambda_2 |f_2(0+iy) | \\
   &     \leq  &  \tr M^{\dg} C^{iy/2} C^{-1/2} \big(\lambda_1 A_1 + \lambda_2 A_2 \big)
                     C^{-1/2} C^{-iy/2} M     \nonumber \\   \label{Mbnd}
    & = &         \tr M^{\dg}  C^{iy/2}      C^{-iy/2}  M  = \tr  M^{\dg} M  
\ee
since  $C $ was defined as $\lambda_1A_1 + \lambda_2   A_2$.  One can 
similarly show that   $|f(1+iy) | \leq \tr  M^{\dg} M  $ which
implies  (\ref{eq:Mbd}).  This establishes  the concavity of
$A \mapsto  \tr A^p K^{\dg} A^{1-p} K$.

The general case then follows from the observation 
\be   \label{eq:trick}
     \tr A^p K^{\dg} B^{1-p} K = 
     \tr  \pmx A & 0 \\ 0 & B \emx^p  \pmx 0 & K^{\dg} \\ 0 & 0 \emx 
         \pmx A & 0 \\ 0 & B \emx^{1-p}  \pmx 0 & 0 \\ K  & 0 \emx .     
         \ee

\noindent{\bf Extension to infinite dimensions:}  The restriction to finite 
dimensional matrices was used only to
ensure that $C^{-1}$ is bounded on the orthogonal complement of $\ker(C)$ so
 that \eqref{unifbnd} gives a uniform bound for $| f_k(z)|$ 
 on the  strip $0 \leq  {\rm Re} \,(z) \leq 1$.    It is worth
 emphasizing that in this part of the proof it is enough to
 show that $| f_k(z)|$ satisfies {\em some} upper bound, which can be 
 rather crude as long as it holds uniformly for all $z$ in the infinite strip.
 Only  for the subsequent estimate on the boundary do we need
a precise  bound of the form \eqref{Mbnd},  which can be generalized
 to operators on infinite dimensional spaces.
    
  Therefore, the theorem can be extended to infinite dimensions 
in several ways.     First, observe that $C = \lambda_1 A_1 + \lambda_2 A_2$
 implies that for $0 \leq q \leq 1$ 
\be
   A_k \leq \lambda_k^{-1} C ~ \Raw~ A_k^q \leq \lambda_k^{-q} C^q
     ~ \Raw~ C^{-q/2} A_k^q C^{-q/2} \leq \lambda_k^{-q} I
\ee
where the first implication uses the operator monotonicity of the
map $A \raw A^q$.   Then
  under the additional hypothesis that $K$ is Hilbert-Schmidt 
  (i.e., $\tr K^{\dg} K < \infty$),    
    one can  replace  (\ref{unifbnd}) by 
\be
|f_k(z)| \leq  \lambda_k^{-1}  \, \tr M^{\dg} M \leq  
            \lambda_k^{-1}  \,  \|C \| \, \tr K^{\dg} K.
  \ee

Lieb uses the even weaker assumption that $M = C^{q/2} K C^{p/2}$ is
Hilbert-Schmidt, and also  proves concavity for the map
$(A,B) \mapsto  \tr A^p K^{\dg} B^q K$ for $0 \leq p+q \leq 1$.  For this,
he uses the operator concavity of the map  $A \mapsto A^t$ for $0 \leq t \leq 1$
to conclude that  $C^{-t/2} \big(\lambda_1A_1^t + \lambda_2 A_2^t \big) C^{-t/2} \leq I$.

\section{Connecting the concavity theorems}  \label{section:chain}

\subsection{Non-commutative multiplication and differentiation}  \label{sect:omega}

The key to connecting the two concavity results mentioned above 
was Lieb's 
  realization that, for any fixed positive semi-definite matrix $A$,
 the following two linear maps on $n \times n$
matrices are inverses of each other, i.e.,
\be
  X   = \Omega_A(K) &  \equiv  & \int_0^1 A^p K A^{1-p} dp \\
   \iff  \hskip3cm & ~ &    \nonumber \\  
      K =  \Omega_A^{-1}(X)&  \equiv  & \int_0^{\infty} \frac{1}{A+ uI} X \frac{1}{A+ uI} du .
\ee
This is far from obvious, but can be verified by expanding in a basis of
eigenvectors of $A$.    Moreover, $ \Omega_A$ can be regarded as
a non-commutative version of multiplication by $A$
and $ \Omega_A^{-1}$ as a non-commutative version of multiplication 
by $A^{-1}$.     Both operators are positive semi-definite
with respect to the inner product $\tr A^{\dg} B = \bra A, B \ket$, i.e.,
$\tr K^{\dg}  \Omega_A(K) \geq 0$ and $\tr X^{\dg}  \Omega_A^{-1}(X) \geq 0$.

The operator  $\Omega_A^{-1}$ arises when one uses the integral representation
\be   \label{eq:intrep1}
   \log P - \log Q  & = &  \int_0^{\infty}  \Big[ \frac{1}{Q+ uI} - \frac{1}{P+ uI} \Big] du \\
      & = &   \int_0^{\infty}  \Big[ \frac{1}{Q+ uI}  (P - Q)  \frac{1}{P+ uI} \Big] du    \label{eq:intrep2}
\ee
to compute derivatives which arise in studying entropy.    
In particular,  when $f(x) = \log(A + x K)$ with
$K = K^{\dg}$ self-adjoint,  it follows from (\ref{eq:intrep1}) that
$f^{\prime}(0) = \Omega_A^{-1}(K)$  and 
$f^{\prime \prime}(0) = - 2 \Upsilon_A(K) $ where
\be
   \Upsilon_A(K) =  \int_0^{\infty} 
              \frac{1}{A+ uI} K^{\dg} \frac{1}{A+ uI}  K \frac{1}{A+ uI} du .
\ee 
This leads  \cite{OP} to the useful (and norm convergent) expansion
\be   \label{eq:logseries}
  \log(A + x K) = \log(A) + x  \Omega_A^{-1}(K) - x^2   \Upsilon_A(K) + \ldots
\ee

\subsection{From Theorem~\ref{thm1} to Theorem~\ref{thm2} } 

   To show that Theorem~\ref{thm1} implies Theorem~\ref{thm2}, first observe
   that the joint convexity of $X^{\dg}\Omega_A^{-1}(X)$ is equivalent to the
   inequality
   \be   \label{eq:xy1}
 \lefteqn{  s  \tr X^{\dg} \Omega_A^{-1}(X) +  (1 \mm s) \tr Y^{\dg} \Omega_B^{-1}(Y)  } \\  
 &   \geq  &  \tr (s X + (1-s) Y)^{\dg}  \Omega_{sA+ (1 \mm s)B}^{-1}(s X + (1 \mm s) Y)  \nn
   \ee  
which can be rewritten as   
\bsq  \label{eq:xy2} \be   \label{eq:xy2a}
\lefteqn{    \tr \pmx X^{\dg} & Y^{\dg} \emx 
       \pmx    s \Omega_A^{-1} & 0 \\ 0 &  (1 \mm s) \Omega_B^{-1} \emx    \pmx X \\ Y \emx  } \\
    &   \geq  &   \pmx X^{\dg} & Y^{\dg} \emx   \Omega_{sA+ (1 \mm s)B}^{-1}   \ot 
        \pmx s^2 & s (1 \mm s) \\  s (1 \mm s) & (1 \mm s)^2 \emx 
          \pmx X \\ Y \emx   .    \label{eq:xy2b}
    \ee \esq
We now regard $A,B, s$ as fixed and consider the ratio obtained 
by dividing   (\ref{eq:xy2b})  by (\ref{eq:xy2a}).  
Joint convexity holds if this ratio is $\leq 1$ for all choices of $(X, Y)$.       
 This is an optimization problem of
   the form   $ \sup_v \dfrac{ \bra v, C v \ket} { \bra v, D v \ket} $ 
   which is    equivalent to the  eigenvalue problem  $Cx = \lambda  Dx $ when 
   $C, D$ are positive   semi-definite.     For the operators used here, the reduction yields
   a pair   of equations
\bsq \label{eq:evalpair}     \be 
  \Omega_{sA+ (1 \mm s)B }^{-1}(sX + (1 \mm s)Y) & = & \lambda  \Omega_A^{-1}(X) \\
 \Omega_{sA+ (1 \mm s)B }^{-1}(sX + (1 \mm s)Y) & = & \lambda  \Omega_B^{-1}(Y) 
 \ee \esq
and we want to show that $\lambda \leq 1$.    Define
$M =      \Omega_{sA+ (1 \mm s)B }^{-1}(sX + (1 \mm s)Y)$ and observe that
(\ref{eq:evalpair} ) gives three expressions for   $M$ since
\be
   M = \Omega_{sA+ (1 \mm s)B }^{-1}(sX + (1 \mm s)Y) = \lambda  \Omega_A^{-1}(X)
     =   \lambda  \Omega_B^{-1}(Y) .
\ee
Applying the    appropriate inverse operator to each of these
and combining the last two, one finds
\bsq \label{eq:invevalpair}     \be 
sX + (1 \mm s)Y & = &   \Omega_{sA+ (1 \mm s)B }(M) \\
\lambda [sX + (1 \mm s)Y] & = &  s \Omega_A(M) +  (1 \mm s)\Omega_B(M) 
\ee \esq   
Now Theorem~\ref{thm1} implies 
\be
s \tr M^{\dg} \Omega_A(M) +  (1 \mm s) \tr M^{\dg} \Omega_B(M) \leq   
  \tr M^{\dg}  \Omega_{sA+ (1 \mm s)B }(M),
\ee
which then implies
\be
   (1 - \lambda) \tr M^{\dg} [sX + (1 \mm s)Y]  
     =  (1 - \lambda) \tr M^{\dg} \Omega_{sA+ (1 \mm s)B }(M) \leq 0 .
\ee
It then follows from the fact that $ \Omega_{sA+ (1 \mm s)B }$ is
positive semi-definite that $\lambda \leq 1$.   
    
   \subsection{From Theorem~\ref{thm2} to Theorem~\ref{thm3} }  \label{imp:23}

Showing that Theorem~\ref{thm2}  implies the concavity of the map
 $A \mapsto \tr e^{K + \log A}$ is elementary, but tedious, as it 
 is ``merely'' a matter of computing derivatives with attention
 to non-commutativity.    Let $f(x) = \tr e^{K + \log(A + xB)}$.
 Then $f^{\prime \prime}(0) \leq 0$ implies that $\tr e^{K + \log A}$
 is concave.   One can use power series to verify that  
  $  \frac{d~}{dx} e^{F+xG} = \Omega_{e^F}(G)$.
This can then be combined with (\ref{eq:logseries}) to yield
\be   \label{eq:deriv}
   f^{\prime \prime}(0) & = &
       \tr  \Omega_A^{-1 }(B) \Omega_{e^{K + \log A}}\big[ \Omega_A^{-1 }(B)\big]  
       - 2 \tr e^{K + \log A}  \Upsilon_A(B).   
    \ee
(Chapter 3 of \cite{OP} gives a clear exposition of the details.) To show   
 that  (\ref{eq:deriv}) is negative,
one can apply the inequality (\ref{eq:hom_deriv}) to $g(x) = G(A+xD,K+xL)$
with $G(A,K) = - \tr K \Omega_A^{-1}(K)  $. 
One finds
\be  \label{eq:comp.deriv}
 g^{\prime}(0) & = &  - 2 \tr D  \Upsilon_A(K)
     + 2 \tr L \Omega_A^{-1}(K)          \leq  \tr L \Omega_D^{-1}(L).
\ee
Now choose $K =   B , D = e^{K + \log A}$ and 
   $L =   \Omega_{e^{K + \log A}}\big[ \Omega_A^{-1 }(B)\big] $.  
Then    $  \tr L \Omega_D^{-1}(L) =  \tr L  \Omega_A^{-1 }(B) $.
Making the appropriate substitutions in (\ref{eq:deriv}) and
and (\ref{eq:comp.deriv}) yields
\be
   f^{\prime \prime}(0) \leq   2  \Big(  \tr L \Omega_D^{-1}(L)
 -   \tr L \Omega_A^{-1}(B) \Big) = 0.
\ee
Although we have sketched the path from concavity of $\tr K^{\dg} A^p K A^{1-p}$
to concavity of $\tr e^{K + \log A}$, most of the steps are easily seen to be
reversible.   With a bit more effort \cite{Lb}, one can show that the reverse implication
also holds.

  \section{Theorem~\ref{thm2}: a direct proof and  implications} \label{sect:newpf}
  
  \subsection{Proof} 
Lieb and Ruskai \cite{LbR2} proved the joint
operator convexity of the map  $(X,A) \mapsto  X^{\dg} A^{-1} X$.
Unfortunately,  this does not seem to directly imply the joint convexity of  
$\tr X^{\dg} \Omega_A^{-1}(X) $.      However, the following observation
allows one to adapt the argument in \cite{LbR2} to give another proof of
the  joint convexity of  $\tr X^{\dg} \Omega_A^{-1}(X) $.   Let $L_A$ and
$R_A$ denote left and right multiplication by $A$ so that
$L_A(X) = AX$ and $R_A(X) = XA$.    By expanding in a basis in which
$A$ is diagonal, one can verify that
\be   \label{intrep2}
  \tr K^{\dg} \Omega_A^{-1}(K)   =  
        \tr \int_0^{\infty}  K^{\dg} (L_A + t R_A)^{-1}(K) \frac{1}{1+t} dt,
 \ee
and
\be  \label{intrep.ups}
  \tr A \Upsilon_A(K) = 
   \tr \int_0^{\infty}  K^{\dg} (L_A + t R_A)^{-1}(K) \frac{1}{(1+t)^2} dt.
\ee
The operator $(L_A + t R_A)^{-1}$ can be regarded as another
non-commutative version of multiplication by $A^{-1}$.  It is
positive semi-definite with respect to the Hilbert-Schmidt inner
product since both $L_A$ and $R_A$  are self-adjoint and
positive semi-definite.  For example,  $\tr [L_A(W)]^{\dg} X =
\tr (AW)^{\dg} X = \tr W^{\dg} AX = \tr W^{\dg} L_A(X)$ and
$ \tr X^{\dg} L_A(X) \geq 0$.

\begin{lemma} \label{lemma.Schwz}
For each fixed $t \geq 0$ the map $(A,K) \mapsto  \tr K^{\dag} (L_{A } + t R_{A })^{-1}(K)$
is jointly convex.
\end{lemma}
\noindent{\bf Proof:}
Let  $M_j = (L_{A_j} + t R_{A_j})^{-1/2}(K_j) -  (L_{A_j} + t R_{A_j})^{1/2}(\Lambda) $.
Then
\be  \label{eq:Schz1}
   0  & \leq & \sum_j \tr M_j^{\dg} M_j   \nonumber  \\   
     & = &    \sum_j \tr K_j^{\dg}  (L_{A_j}  + t R_{A_j})^{-1}(K_j)  - 
     \tr   \big( \ts{ \sum_j } K_j^{\dg}  \big) \Lambda    
               \\    & ~  & ~ \qquad \qquad  -  \tr \Lambda^{\dg} \big( \ts{\sum_j K_j } \big)
        +    \tr \Lambda^{\dg}    \ts{ \sum_j} \big( L_{A_j}  + t L_{R_j}) \Lambda  .   \nonumber 
         \ee
Next, note that  
$ \sum_j  \big( L_{A_j} + t R_{A_j})(W) =   \sum_j \big( A_j W + t W A_j \big) 
=   \big( \ts{ \sum_j} A_j \big) W + t W   \big( \ts{ \sum_j} A_j \big)    \linebreak       
                = L_{\sum_j A_j}(W)  +  t R_{\sum_j A_j}(W)  $ for any matrix $W$.
Therefore, inserting the choice
 $\Lambda =   \big(  L_{\sum_j A_j}   +  t R_{\sum_j A_j} \big)^{-1}  
         \big( \ts{\sum_j K_j } \big) $
in  (\ref{eq:Schz1}) yields
 \be    \label{eq:Schwzt}
  \tr   \big( \ts{ \sum_j } K_j \big)^{\dg}   \big(  L_{\sum_j A_j}   +  t R_{\sum_j A_j} \big)^{-1}  
         \big( \ts{\sum_j K_j } \big)
  \leq   \sum_j \tr K_j^{\dg}  (L_{A_j} + t R_{A_j})^{-1}(K_j) .
            \ee
for any $t \geq 0$.     The joint convexity then follows from the replacement
$A \raw \lambda_j A$, $ K \raw \lambda_j K$ with $\lambda_j > 0$ and $ \sum_j \lambda_j = 1$.
Alternatively, one can simply observe that 
$\tr \lambda K^{\dag} (L_{\lambda A } + t R_{\lambda A })^{-1}(\lambda K) 
   = \lambda \tr K^{\dag} (L_{A } + t R_{A })^{-1}(K)$ and use the first observation
 in Appendix~\ref{app:homo} to reduce the joint convexity to  (\ref{eq:Schwzt}).
  \qquad {\bf QED} 

The joint convexity of both $ \tr K^{\dg} \Omega_{A}^{-1}(K) $ and
$ \tr A \Upsilon_{A}^{-1}(K) $ follow immediately from Lemma~\ref{lemma.Schwz}
and the integral representations above.  In particular,
inserting (\ref{eq:Schwzt}) in  (\ref{intrep2}) yields
\be
   \tr  \big( \ts{\sum_j K_j } \big)^{\dg} \Omega_{\sum_j A_j}^{-1} \big( \ts{\sum_j K_j } \big) 
    \leq  \sum_j   \tr K_j^{\dg} \Omega_{A_j}^{-1}(K_j)
\ee      
which proves Theorem~\ref{thm2} when $A = B$. 
The general case then follows as in (\ref{eq:trick}).
The choice $t = 0$ in Lemma~\ref{lemma.Schwz} yields the operator
convexity of $X^{\dg} A^{-1} X$ proved in \cite{LbR2}.

  \subsection{From Theorem~\ref{thm2} to  entropy inequalites}
  The von Neumann entropy 
is defined as $S(P) = - \tr P \log P$ and the relative entropy as
 $H(P,Q) \equiv \tr P \big( \log P - \log Q \big)$ with $P,Q$ 
 positive semi-definite and $\ker(Q) \subset \ker(P)$. 
 (One usually assumes that $\tr P = \tr Q = 1$, but this is
 not strictly necessary.)    Several special cases of the joint
 convexity of  $H(P,Q) $ follow immediately from Theorem~\ref{thm2}.

 Using (\ref{eq:logseries}), one can show that
$\frac{d~}{dx} S(A + x K) \big|_{x = 0} = - \tr K \log(A)$, when $\tr K = 0$
and  that the second derivative satisfies
\be    \label{eq:delta} 
 - \Delta(A,K) &  \equiv  & \tfrac{d^2~}{dx^2} S(A + x K) \big|_{x = 0} \\ 
 & = &     - \tr K \Omega_A^{-1}(K) + \tr A  \Upsilon_A(K) \\
       & = & - \tr \int_0^{\infty}  K^{\dg} (L_A + t R_A)^{-1}(K) \frac{t}{(1+t)^2} dt   \label{eq:delta}
       \ee 
where we used the integral representations   (\ref{intrep2}) and (\ref{intrep.ups})
together with the simple fact $\frac{1}{1+t} - \frac{1}{(1+t)^2} = \frac{t}{(1+t)^2}$.
Lemma~\ref{lemma.Schwz} implies that   $\Delta(A,K)$ is  jointly convex.
One also has
 \be
    \frac{ \partial^2}{\partial a \partial b} H(P + a A, P + b B) \Big|_{a = b = 0}=
       \tr A^{\dg} \Omega_P^{-1}(B).
 \ee

Now consider density matrices defined on a tensor product
of two or three spaces.   First, suppose that 
$\gamma_{12}  = \pmx A & 0 \\ 0 & B \emx$ is block diagonal
with $A, B$ in ${\mathcal H}_2$ so that $S(\gamma_{12}) = S(A) + S(B)$
and $\gamma_2 = \tr_1 \gamma_{12} = A + B$.  Then
 \be
   S(\gamma_{12}) - S(\gamma_2) = S(A) + S(B) - S(A + B)
 \ee 
Now consider the perturbed density matrix
$\gamma_{12}  = \pmx A + x K & 0 \\ 0 & B + x L \emx$
and let $g(x) = S(A+xK) + S(B+xL) - S[A + B + x (K + L)]$.
Then
  \be
g^{\prime \prime}(0) =
        - \Delta(A,K) - \Delta(B,L) + \Delta(A+B,K + L) \leq 0
\ee  
because $ \Delta(A,K)$ is jointly convex in $(A,K)$.   This
proves the concavity of the map 
$\gamma_{12} \mapsto S(\gamma_{12}) - S(\gamma_2)$ in the
special case of block diagonal matrices.

 The situation above can be compared to
 $\rho_{123}  = \pmx t \rho_{12}^{\prime} & 0 \\ 0 & (1-t) \rho_{12}^{\prime\prime} \emx$. 
In this case, the strong subadditivity inequality in the form
$S(\rho_{123}) - S(\rho_{23}) \leq  S(\rho_{12}) - S(\rho_2)$
is precisely the  concavity of $\rho_{12} \mapsto S(\rho_{12}) - S(\rho_2) $
with $\rho_{12} = t \rho_{12}^{\prime} + (1-t) \rho_{12}^{\prime\prime} $ 
  arbitrary.
 In the previous paragraph, $\rho_{12}^{\prime}$ and $\rho_{12}^{\prime \prime}$
 were also required to be block diagonal.

The joint convexity of the less familiar symmetrized relative entropy
follows immediately from the joint convexity of $ \tr X^{\dg} \Omega_A^{-1}(X)$.
Using (\ref{intrep2})  one easily finds 
 \be
  H(P,Q) + H(Q,P) =
  \tr  \int_0^{\infty}  \Big[(P - Q)  \frac{1}{Q+ uI}  (P - Q)  \frac{1}{P+ uI} \Big] du .
 \ee
It then follows immediately  from Theorem~\ref{thm2} with $K = P - Q$
that $H(P,Q) + H(Q,P)$ is jointly convex in $P,Q$.   This implies that
at least one of $H(P,Q)$ and $H(Q,P)$ is jointly convex in $P,Q$.

Although Theorem~\ref{thm1}  gives a short route to special cases 
of the joint convexity
of relative entropy, proving the general case seems to require one of
the paths through Theorem~\ref{thm1} or Theorem~\ref{thm3}. 
Bauman  \cite{B} also recognized that there is a connection between the
concavity of $S(\rho_{12}) - S(\rho_2)$  for block diagonal $\rho_{12}$
and joint convexity of 
$ \tr X^{\dg} \Omega_A^{-1}(X)$ for self-adjoint $X$;
and proved the latter for  $2 \times 2$ matrices.

\subsection{Further remarks}

The argument used to prove Lemma~\ref{lemma.Schwz} is remarkably 
simple\footnote{In quant-ph/0604206 the 
argument presented in Section 4.1 is used to give a direct proof of the
joint convexity of relative entropy.} 
(and can be applied  to $ L_{A_j} + t R_{B_j}$ to give a direct proof of the general case
 of Theorem~\ref{thm2}.)    However, the insight needed to
rewrite $ \tr K^{\dg} \Omega_A^{-1}(K)$ in the form (\ref{intrep2}) comes from
the very powerful relative modular operator formalism originally developed
by Araki \cite{Ak} to extend relative entropy to infinite dimensional operator
algebras which might not even have a trace.    The utility of this formalism in
finite dimensional situations was not realized until much later.   For an
exposition see Ohya and Petz \cite{OP}.  Lemma~\ref{lemma.Schwz} can
also be used to prove the joint convexity of any operator which has an
integral representation as in (\ref{intrep2}) or (\ref{intrep.ups})  
with a positive weight, and the proof
presented here is similar to one presented in \cite{LesR} to prove
monotonicity of a generalized form of relative entropy and related inequalities 
within the theory of monotone Riemannian metrics developed by Petz \cite{Pz}.

In view of connections to the Bures metric and quantum channels, we make a few further
 remarks.   Monotone means decreasing
under completely positive trace preserving maps,
and  (\ref{intrep2}) is the Riemannian metric associated
with the usual relative entropy $H(P,Q)$.
A large class of such metrics,  generalizations of relative entropy, and of
the operator $\Omega_A^{-1}$
can be obtained by replaced the weight $\frac{1}{1+t} $ in (\ref{intrep2})
by one coming from a subclass of monotone operator functions. 
The bound,
\be
    R_A^{-1} +  L_A^{-1}   \geq  \Omega_A^{-1} \geq  (R_A + L_A)^{-1},
\ee
which holds as an operator inequality is also satisfied by these
generalizations of $\Omega_A^{-1}$, and the upper and lower
bounds are special cases of these generalizations.    This is of
interest because any of these Riemannian metrics can be
used to define a geodesic distance  $D(P,Q)$ between two
density matrices.   The Bures metric given by
$[D_{\rm Bures}(P,Q) ]^2 = 2 \big[ 1 - \tr \big(\sqrt{P} Q \sqrt{P} \big)^{1/2} \big]$
is precisely the geodesic distance associated with
$\tr X^{\dg} (R_A + L_A)^{-1}(X)$.    The importance of the Bures
metric, which is closely related to the fideity, in quantum information
has been emphasized by Uhlmann  \cite{Uhl2}.      Unfortunately, no
closed form expression for the geodesic distance associated with
(\ref{intrep2}) is known.    See \cite{G,Jc} for bounds
and a discussion of the geodesic distances for the 
Wigner-Yanase-Dyson entropies, which are closely related to the
function considered in Theorem~\ref{thm1}.

\noindent{\bf Acknowledgment:}   It is a pleasure to thank Professor M. d'Ariano for 
 the opportunity  to present lectures during the Quantum Information Processing
 workshop in Pavia, Italy.

\appendix 

\section{Maximum modulus principle}  \label{app:maxmod}

The maximum modulus principle given in most elementary texts,
states that the modulus of an analytic function can not have
a local maximum on a bounded open set unless it is a constant. 
It then follows that a function $f(z)$ which is analytic on  a bounded set
and continuous on its closure achieves the supremum of $|f(z)|$
on the boundary.   

Now consider a function $f(z)$ which is analytic on the strip $\{ z : 0 < {\rm Re} \, z < 1 \}$
and continuous and uniformly bounded on its closure 
$\{ z : 0 \leq {\rm Re}\,  z \leq 1 \}$.
Let $M = \sup \{ |f(z)| :    0 \leq {\rm Re} \, z \leq 1 \}$.    The maximum
modulus principle for a strip \cite{G} says that $M$ is equal to the supremum 
of $|f(z)|$ on the boundary, i.e., 
$M =  \sup \{ |f(z)| :  z= 0 + i y ~\text{or} ~ z = 1 + iy  \}$.

If the supremum is actually attained at some point $\wtd{z} = a + i y$,
i.e.,  $|f(\wtd{z})| = M$, the result follows easily from the theorem
for bounded regions.     The possibility of  $a \neq 0,1$ can be excluded
by considering $f(z)$ on the rectangular
 region  ${\mathcal Q}_b = \{ z : 0 \leq {\rm Re} Z \leq 1 ~\text{and}~
 -b \leq  {\rm Im} z \leq b \}$ with $b > y$.   Since $f$ cannot have a 
 relative maximum on    ${\mathcal Q}_b $, one has a contradiction
 unless $f$ constant.  If $a = 0$ or $a = 1$,
then $\wtd{z}$ lies on the boundary of the strip and the assertion holds.   
To prove the result if 
the supremum is not attained, consider the functions
$f_{\varepsilon}(z) \equiv z^{-\varepsilon} f(z)$.    Since
$\ds{\lim_{y \raw \pm \infty} |f_{\varepsilon}(x + i y)| = 0}$, the maximum modulus
$M_{\varepsilon} = \sup \{ |f_{\varepsilon}(z)| :    0 \leq {\rm Re} \, z \leq 1 \}$
is attained for some $z_{\varepsilon}  = iy$ or $z_{\varepsilon} = 1 + iy$.   But
$|f_{\varepsilon}(z)| \leq |f(z)|$, and $M = \ds{\lim_{\alpha \raw 0}} M_{\varepsilon} = 
  \sup \{ |f(z)| :  z= 0 + i y ~\text{or} ~ z = 1 + iy  \}$.

 \section{Homogenous concave functions}  \label{app:homo}
 
 Most of the functions considered here are homogenous of
 degree one, i.e., $g(\mu A) = \mu \, g(A)$.   In this case concavity
 is equivalent to superadditivity.   To see this observe that
homogeneity and concavity at $\half$ imply
 \be
     \half g(A+B) =  g\big(\half[A+B] \big) \geq \half g(A) + \half g(B) .
 \ee
 Conversely, if $g$ is homogenous and superadditive
 \be
     g\big[x A + (1-x) B\big] \geq  g(xA) + g[(1-x)B] = x g(A) + (1-x) g(B)
 \ee

One extremely useful property of homogenous concave functions
is the following derivative inequality \cite{R}.
\be  \label{eq:hom_deriv}
    \lim_{x \raw 0}  \frac{g(A + x B) - g(A) }{x}  \leq g(B)   
     \ee
which follows easily from $g(A + x B) \leq  g(A) + x g(B)$.


\section{Erratum to quant-ph/0404126 by M.B. Ruskai} 

Since this note appeared, people who have presented the proof in class
have reported some minor errors and omissions in the argument presented in the note. 
The problems and necessary modifications are described in detail below.

First, the cyclicity of the trace and procedure described above   \eqref{unifbnd}  yield 
\be  \label{newbd}
  |f_k(z))|  \, \leq \,   \| A_k \|^{1-x} \, \| C^{-1} \|^{1-x} \, 
      \tr | M C^{-(x+iy)/2} A_k^{x+iy} C^{-(x+iy)/2} M^\dag |  ~~.
\ee
For $y \neq 0$ one can {\em not} remove  the $ | ~ | $ 
because the quantity inside is not positive semi-definite.   
From a pedagogical point of view, the simplest argument may
be to let $G(z) = C^{-z/2} A_k^z C^{-z/2}$ and write $f_k(z) =  \tr M^\dag G_k(1-z) M G_k(z)$.
Then 
use \eqref{cs} to obtain
\be
   |f_k(z)| \leq  \big( \tr M^\dag G_k(1-z) G_k(1 - \ovb{z} ) M \big)^{1/2}
       \big( \tr M^\dag G_k(\ovb{z}) G_k(z) M \big)^{1/2}.
\ee
Then, the cyclicity of the trace and proceedure described above (5) yields
\be
  \tr M^\dag G_k(\ovb{z}) G_k(z) M & = & \tr  G_k(\ovb{z}) G_k(z) M M^\dag  \nn \\
     & \leq &  \| C^{-1} \|^{2x}  \,  \| A_k \|^{2x} \,  \tr  M^\dag M.
\ee
Combining this with a similar estimate for
$\tr M^\dag G_k(1-z) G_k(1 - \ovb{z} ) M$ and taking the square
root gives the desired result.

Another alternative is to realize that one only needs a bound 
independent  of $z$,
not one with the precise form  \eqref{unifbnd}. 
The polar decomposition theorem implies that one can write any operator
as $X = V |X|$ with $V$ unitary and $|X| = \sqrt{X^\dag X}$ so that $|X| = V^\dag X$.
Using this in \eqref{newbd} yields
\be
  |f_k(z))|   \, \leq \,   \| A_k \| \, \| C^{-1} \|  \, \tr |M^\dag V^\dag M|  
       \, \leq \,    \| A_k \|  \, \| C^{-1} \|  \, \| M \|   \, \tr |M| .
\ee
Although the unitary $V$ may depend on $z$, the last bound on the 
right is  independent of $z$.   One could also use  \eqref{cs} to show
 $\tr |M^\dag V^\dag M|  \leq   \tr M^\dag M \equiv \| M \|_2^2 $; however,
  the weaker bound above  will suffice.
 
Second, to bound \eqref{eq6} one needs   the inequality
\be  \label{cs}
  | \tr X^\dag Y | \, \leq \,   ( \tr X^\dag X  )^{1/2}    \, ( \tr Y^\dag Y  )^{1/2}  \, 
     \leq \,  \half (  \tr X^\dag X +  \tr Y^\dag Y)
\ee
and \eqref{eq7} should be replaced by 
\be
  | f_k(0+iy)| &  \leq &  \half \big( \tr M^{\dg} C^{iy/2} C^{-1/2} A_k C^{-1/2} C^{-iy/2} M    \\  \nn 
     & ~ & \quad + ~ \tr M^{\dg} C^{-iy/2} C^{-1/2} A_k C^{-1/2} C^{iy/2} M \big)   \qquad k=1,2. 
\ee
Then \eqref{eq8} becomes
\bee  
  |f(0+iy) | & \leq  &  \lambda_1|f_1(0+iy) | + \lambda_2 |f_2(0+iy) | \\
   &     \leq  &  \half \big(\tr M^{\dg} C^{iy/2} C^{-1/2} \big(\lambda_1 A_1 + \lambda_2 A_2 \big)
                     C^{-1/2} C^{-iy/2} M     \nonumber \\   \nonumber
                       & ~ & + ~ 
                        \tr M^{\dg} C^{-iy/2} C^{-1/2} \big(\lambda_1 A_1 + \lambda_2 A_2 \big)
                     C^{-1/2} C^{+iy/2} M  \big)  \\
    & = &    \half \big(     \tr M^{\dg}  C^{iy/2}      C^{-iy/2}  M +   \tr M^{\dg}  C^{-iy/2}      C^{+iy/2}  M \big)  
   ~ = ~   \tr  M^{\dg} M.  
\eee

\noindent{\bf Acknowledgment:}  The author is grateful to M. d'Ariano and
P. Hayden for correspondence about these issues.

\end{document}